\documentclass[journal=jacsat,manuscript=article]{achemso}

\usepackage[version=3]{mhchem} 
\usepackage{comment}
\usepackage[hidelinks]{hyperref}
\usepackage{xcolor}
\usepackage{siunitx}
\usepackage{bm}
\usepackage{amsmath,amssymb}
\usepackage{xr}
\usepackage{nicefrac}

\DeclareUnicodeCharacter{03C0}{\ensuremath{\pi}}
\newcommand{\didv}{{d}\textit{I}/{d}\textit{V} }

\author{Gon\c{c}alo Catarina}
\email{goncalo.catarina@empa.ch}
\author{Elia Turco}
\author{Nils Krane}
\author{Max Bommert}
\author{Andres Ortega-Guerrero}
\author{Oliver Gr\"{o}ning}
\author{Pascal Ruffieux}
\email{pascal.ruffieux@empa.ch}
\author{Roman Fasel}
\affiliation[Empa]{nanotech@surfaces Laboratory, Empa---Swiss Federal Laboratories for Materials Science and Technology, 8600 D\"{u}bendorf, Switzerland}
\alsoaffiliation[University of Bern]{Department of Chemistry, Biochemistry and Pharmaceutical Sciences, University of Bern, 3012 Bern, Switzerland}
\author{Carlo A. Pignedoli}
\email{carlo.pignedoli@empa.ch}
\affiliation[Empa]{nanotech@surfaces Laboratory, Empa---Swiss Federal Laboratories for Materials Science and Technology, 8600 D\"{u}bendorf, Switzerland}

\title{Conformational tuning of magnetic interactions in coupled nanographenes}

\begin{document}

\begin{abstract} 
Phenalenyl (\ce{C13H9}) is an open-shell spin-$\nicefrac{1}{2}$ nanographene.
Using scanning tunneling microscopy (STM) inelastic electron tunneling spectroscopy (IETS), covalently-bonded phenalenyl dimers have been shown to feature conductance steps associated with singlet-triplet excitations of a spin-$\nicefrac{1}{2}$ dimer with antiferromagnetic exchange. 
Here, we address the possibility of tuning the magnitude of the exchange interactions by varying the dihedral angle between the two molecules within a dimer. 
Theoretical methods, ranging from density functional theory calculations to many-body model Hamiltonians solved within different levels of approximation, are used to explain STM-IETS measurements of twisted phenalenyl dimers on a h-BN/Rh(111) surface. 
By means of first-principles calculations, we also propose strategies to induce sizable twist angles in surface-adsorbed phenalenyl dimers via functional groups, including a photoswitchable scheme.
This work paves the way toward tuning magnetic couplings in carbon-based spin chains and two-dimensional lattices.
\end{abstract}


\section{Introduction} 

Atomically precise graphene nanostructures, known as nanographenes, have emerged as promising materials for electronic, optic, and magnetic applications\cite{gu_nanographenes_2022}.
A remarkable example is the case of graphene nanoribbons\cite{cai_atomically_2010}, whose geometry-dependent bandgap\cite{son_energy_2006,chen_tuning_2013}
and topology\cite{cao_topological_2017,groning_engineering_2018}
hold potential for integration in electronic devices\cite{wang_graphene_2021}.
Additionally, advances in on-surface synthesis have ignited the field of carbon-based magnetism\cite{oteyza_carbon-based_2022},
enabling the fabrication of a variety of open-shell nanographene molecules, such as $[n]$-triangulenes\cite{turco_observation_2023,pavlicek_synthesis_2017,mishra_synthesis_2019,su_atomically_2019,mishra_synthesis_2021} (with $n=2,3,4,5,7$), Clar's goblet\cite{mishra_topological_2020}, and [5]-rhombene\cite{mishra_large_2021}.

In recent years, concerted efforts across various disciplines have promoted significant progress in the field of magnetic nanographenes.
With insights from theory\cite{fernandez-rossier_theory_2009,ternes_spectroscopic_2009,ternes_spin_2015},
the open-shell character of nanographenes has been demonstrated through inelastic electron tunneling spectroscopy (IETS) with a scanning tunneling microscopy (STM) tip.
This has enabled the observation of, e.g., spin excitations in [3]-triangulene dimers\cite{mishra_collective_2020} and zero-bias Kondo resonances in 
[2]-triangulenes\cite{turco_observation_2023}.
Furthermore, rationally designed precursors have facilitated the on-surface synthesis of diverse nanographene spin systems, ranging from small clusters\cite{mishra_collective_2020,su_-surface_2021,hieulle_-surface_2021,mishra_nonbenzenoid_2022,cheng_-surface_2022,du_orbital-symmetry_2023,krane_exchange_2023,turco_magnetic_2024}
to one-dimensional chains\cite{mishra_observation_2021,zhao_tunable_2024} 
and two-dimensional lattices\cite{delgado_evidence_2023,catarina_broken-symmetry_2023,frezza_-surface_2024}.

To fully harness the exciting opportunities presented by open-shell nanographenes, it is crucial to gain control over their magnetic exchange couplings.
For instance, in the realm of quantum simulation\cite{georgescu_quantum_2014}, nanographene spin systems with tunable magnetic interactions hold promise for two main reasons.
First, nanographene spin chains have already been shown to realize interacting model Hamiltonians that feature exotic physics such as spin fractionalization\cite{mishra_observation_2021}.
Second, $\pi$-magnetism allows for larger couplings than those of alternative platforms like cold atoms\cite{mazurenko_cold-atom_2017,sompet_realizing_2022}, quantum dots\cite{mortemousque_coherent_2021,kiczynski_engineering_2022},
or atomic spins on surfaces\cite{choi_colloquium_2019,wang_realizing_nodate}.
Noteworthy results have been achieved in the pursuit of tuning exchange interactions in nanographenes.
Precise spin manipulation has been accomplished via STM in Clar's goblet chains\cite{zhao_tailoring_2024,zhao_tunable_2024}, where controlled passivation of spin sites was obtained via tip-induced dehydrogenation.
Tip proximity has been found to alter the magnetic ground state of non-planar diradical nanographenes with anchoring end groups, presumably due to structural changes\cite{vegliante_tuning_2024}.
Nitrogen substitution\cite{wang_aza-triangulene_2022,zhu_collective_2023,yu_predicting_2023,henriques_beyond_2023},
influence of surface\cite{wang_aza-triangulene_2022,krane_exchange_2023,mishra_bistability_2024}, and designer molecular structure\cite{biswas_steering_2023,jacobse_five-membered_2023,fernandez-rossier_magnetism_2007,ortiz_magnetic_2023,sandoval-salinas_electronic_2023,henriques_designer_2024}
have been investigated for their impact on the magnetic properties.

In this work, we explore the potential to manipulate magnetic interactions based on the conformation of covalently-coupled nanographenes.
Specifically, we consider [2]-triangulene (also known as phenalenyl) dimers and study how the exchange interaction $J$ varies with the dihedral angle $\theta$ between the two molecules (Fig.~\ref{fig1}c).
This is motivated by our experimental data obtained from phenalenyl dimers on a h-BN/Rh(111) surface, which show signatures of a twist between the two phenalenyl units, supported by our theoretical analysis.
Nevertheless, the theoretical methodologies presented herein extend beyond this specific case and are particularly relevant for non-planar nanographene spin systems.
Moreover, we propose designer phenalenyl dimers---with anchoring functional groups---that, according to first-principles calculations, are predicted to display sizable twist angles when adsorbed on the prototypical Au(111) surface.
Among these functional groups, one is based on a photoswitchable compound, offering the potential for light-induced control of the magnetic coupling.

Along this research line, it is worth to recall that planar phenalenyl dimers have been probed by scanning tunneling spectroscopy (STS), revealing characteristics of a spin-$\nicefrac{1}{2}$ Heisenberg dimer with antiferromagnetic exchange couplings of $J=41~\si{\milli\electronvolt}$ on Au(111) and $J=48~\si{\milli\electronvolt}$ on NaCl/Au(111)\cite{krane_exchange_2023}.
Further theoretical understanding of the exchange mechanisms in this system has been provided by \citeauthor{jacob_theory_2022} employing multi-configurational methods to solve extended Hubbard models\cite{jacob_theory_2022}.
Additionally, \citeauthor{yu_magnetic_2023} have studied the magnetic interactions of twisted phenalenyl dimers in gas phase using ab initio methods\cite{yu_magnetic_2023}.

\section{Experiments} 

Here, we report the synthesis and magnetic characterization of phenalenyl dimers on a h-BN/Rh(111) surface\cite{corso_boron_2004}.
A submonolayer coverage of 2H-diphenalenyl precursors was sublimed on an epitaxially grown h-BN/Rh(111) surface kept at room temperature. We find that the molecules exhibit affinity for adsorbing along the pore rims of the superstructure, consistent with previous reports for other flat molecules, such as phthalocyanines\cite{iannuzzi_site-selective_2014,liu_interplay_2015}. By sequential tip-induced atomic manipulations, the additional hydrogens on the two \ce{-CH2-} sites were cleaved and the target magnetic phenalenyl dimer was achieved (see SI-I for details). STM imaging clearly shows the appearance of frontier orbital density lobes and nodal planes after manipulation (Fig.~S1). The successful synthesis of phenalenyl dimers is confirmed by STS through the detection of inelastic conductance steps associated with singlet-triplet spin excitations, indicating that the open-shell $S=0$ ground state (where $S$ denotes the total spin) is retained on h-BN/Rh(111). Compared to the magnetic properties previously reported on Au(111) and NaCl/Au(111)\cite{krane_exchange_2023}, we identify three significant differences. First, the conductance steps associated with the spin excitations are remarkably sharp, owing to the absence of states near the Fermi energy of the h-BN/metal substrate\cite{AUWARTER20191} and the decoupling nature of h-BN. Second, after manipulation, different molecules display distinct STM topographies. Third, molecules with different STM appearances exhibit distinct spin excitation energies.

\begin{figure}[t]
  \centering
  \includegraphics[width=0.8\columnwidth]{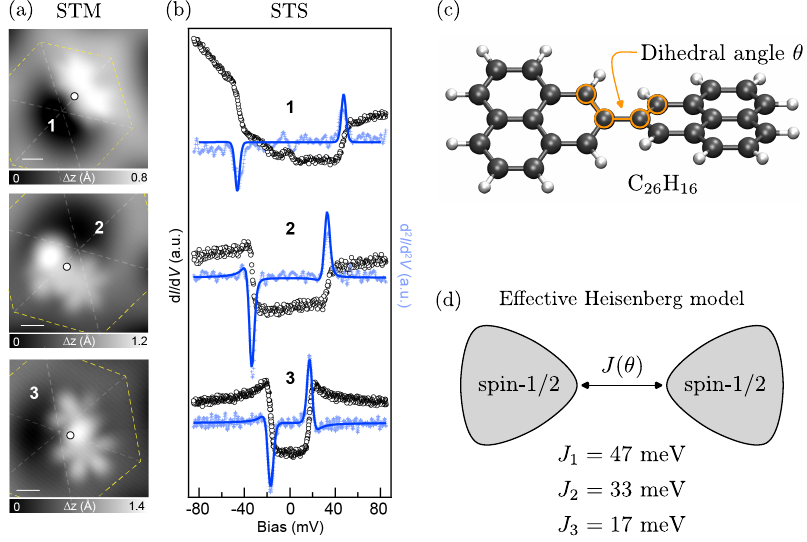}
  \caption{
  (a) STM images of three phenalenyl dimers (\textbf{1}, \textbf{2}, \textbf{3}) adsorbed on a h-BN/Rh(111) surface, featuring different topographies. Yellow dashed lines highlight the hexagonal pore geometry. Tunneling parameters: $V = 1$~V, $I = 10$~pA. Scale bars: 0.5~nm. 
  (b) High-resolution low-bias STS of \textbf{1}, \textbf{2}, and \textbf{3}. Circles in (a) indicate the position where the corresponding \didv spectrum was measured. To extract the spin excitation energy of each \didv spectrum, the corresponding $\mathrm{d}^2I/\mathrm{d}V^2$ spectrum was obtained from numerical differentiation and fitted using a perturbative model\cite{ternes_spin_2015} (blue lines). 
  Open feedback parameters: (\textbf{1}) $V=-100$~mV, $I=150$~pA; (\textbf{2}) $V=100$~mV, $I=100$~pA; (\textbf{3}) $V=100$~mV, $I=250$~pA. Lock-in modulation: $V_{rms}<1$~mV. 
  (c) Ball-and-stick representation of a twisted phenalenyl dimer where $\theta$ is the dihedral angle between the phenalenyls. 
  (d) Effective Heisenberg model for twisted phenalenyl dimers, where $J(\theta)$ denotes the angle-dependent antiferromagnetic exchange coupling between the two spin-$\nicefrac{1}{2}$ phenalenyl units. The exchange couplings indicated correspond to those extracted from the fits shown in (b).
  }
  \label{fig1}
\end{figure} 

Three representative cases are shown in Fig.~\ref{fig1}, where, for each STM topography (panel a), the corresponding low-bias \didv spectrum (where $I$ and $V$ denote the tunneling current and bias voltage, respectively) is presented (panel b). By fitting the $\mathrm{d}^2\textit{I}/\mathrm{d}^2\textit{V}$ spectra\cite{ternes_spin_2015}, we obtain effective antiferromagnetic exchange couplings of 17, 33, and 47~meV.
The molecule labeled as \textbf{1} displays symmetrical orbital density features and the observed $J_1=47$~meV is nearly identical to the value reported for the dimer adsorbed on a flat NaCl/Au(111) surface\cite{krane_exchange_2023}.
For this reason, herein we assume \textbf{1} is planar.
Dimers \textbf{2} and \textbf{3} reveal a pronounced asymmetry in their STM topography, with one phenalenyl unit featuring two and three protruding lobes, respectively, as opposed to the characteristic four-lobed appearance\cite{krane_exchange_2023}. Among the seven dimers analyzed, we observed that all were stably adsorbed on the pore rim, with slight variations in their adsorption configurations. Three of them feature similar STM topography and IETS as \textbf{3} (Fig.~S2), which appears to be the preferred adsorption configuration.

Considering the highly corrugated h-BN/Rh(111) superstructure and the observed differences in STM topography and spin excitation energies, we hypothesize that, due to conformation, the carbon-carbon single bond between the two constituent units of the dimer allows for a twist, i.e., a dihedral angle $\theta$ between the two phenalenyls (Fig.~\ref{fig1}c). We now resort to theory to verify our experimental conjecture and validate the picture of a spin-$\nicefrac{1}{2}$ Heisenberg dimer with $\theta$-dependent antiferromagnetic exchange coupling (Fig.~\ref{fig1}d).

\section{Effective tight-binding model} 

\begin{figure}[t]
 \centering
 \includegraphics[width=0.75\columnwidth]{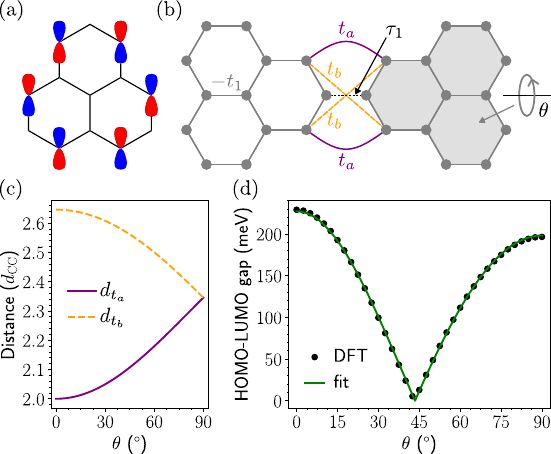}
 \caption{
 (a) Representation of the zero-energy wave function of phenalenyl, as obtained with a nearest-neighbor tight-binding model. 
 Blue and red indicate the sign of the $p_z$ orbitals.
 (b) Effective tight-binding model for twisted phenalenyl dimers.
 $\tau_1$ denotes a nearest-neighbor hopping term that, in contrast to $t_1$, depends on the twist angle $\theta$ between phenalenyls.
 $t_a$ and $t_b$ represent $\theta$-dependent hoppings that allow for hybridization between each phenalenyl's zero-energy state.
 (c) Distance between sites connected by $t_a$ and $t_b$, as a function of $\theta$.
 $d_{\mathrm{CC}}$ is the carbon-carbon distance.
 (d) HOMO-LUMO gap, as a function of $\theta$, obtained with spin-unpolarized DFT (black circles).
 Green line denotes a fit to Eq.~\eqref{eqgapHL}, derived for the model depicted in (b), with Slater-Koster parameterization of $t_a$ and $t_b$.
 }
 \label{fig2}
\end{figure}

We describe nanographenes using a tight-binding model for the $p$ orbitals of the carbon atoms that do not form $sp^2$ hybridization.
Assuming charge neutrality, we consider one electron per orbital.
Within the nearest-neighbor approximation, we find that phenalenyl monomers host one zero-energy state (with significant energy separation from the remaining states\cite{ortiz_theory_2022}), as expected due to their sublattice imbalance\cite{ortiz_exchange_2019}.
As shown in Fig.~\ref{fig2}a, the corresponding wave function is localized at the edge sites of the majority sublattice\cite{ortiz_theory_2022}.
Therefore, in a phenalenyl dimer, despite the lack of sublattice imbalance, the nearest-neighbor tight-binding approximation yields two zero-energy states since first-neighbor hopping does not allow for their intermolecular hybridization.

Intermolecular hybridization of zero-energy states, crucial for the magnetic properties of phenalenyl dimers\cite{jacob_theory_2022,krane_exchange_2023}, occurs if we introduce third-neighbor (or beyond) hopping terms.
In the case of twisted phenalenyl dimers, we consider $t_a$ and $t_b$, as illustrated in Fig.~\ref{fig2}b.
For planar phenalenyls, it is customary\cite{ortiz_theory_2022,jacob_theory_2022} to only account for $t_a$, which corresponds to a third-neighbor hopping.
However, when dealing with twisted dimers, an analysis of the distances (Fig.~\ref{fig2}c) suggests the inclusion of both $t_a$ and $t_b$, as these hopping terms connect equidistant sites for $\theta=90^\circ$.
The presence of intermolecular hybridization results in a splitting between the zero-energy states, which, at the non-interacting level, translates into a gap between the highest occupied molecular orbital (HOMO) and the lowest unoccupied molecular orbital (LUMO).
Using an effective theory\cite{ortiz_theory_2022}, we can derive an analytical expression for the HOMO-LUMO gap,
\begin{equation}
    \Delta_{\mathrm{HOMO-LUMO}}=\frac{2}{3}|t_a-t_b|,
    \label{eqgapHL}
\end{equation}
whose validity is attested in SI-II.

To capture the angle dependence of $t_a$ and $t_b$, we use a two-center Slater-Koster approximation\cite{slater_simplified_1954}, together with the assumption of an exponential decay of the hopping terms with the distance (SI-III).
This approach
is reasonable provided that screening due to non-participant atoms can be disregarded.
However, discrepancies in the literature regarding hopping parameters beyond the standard first-neighbor hopping ($t_1=2.7~\si{\electronvolt}$, which we adopt in this work) pose challenges in estimating the amplitudes and decay rates of $\pi$ and $\sigma$ $p$-$p$ bonds, upon which $t_a$ and $t_b$ rely.
For that matter, we conduct spin-unpolarized density functional theory (DFT) calculations for the HOMO-LUMO gap of twisted phenalenyl dimers (see SI-IV for details of DFT calculations), to which we fit Eq.~\eqref{eqgapHL} (Fig.~\ref{fig2}d).
The agreement between the effective theory and the first-principles calculations, along with the obtained fitting parameters (see SI-III), validates the use of the tight-binding model depicted in Fig.~\ref{fig2}b.
Finally, it must be noted that, while it does not impact the HOMO-LUMO gap, the first-neighbor hopping between the two phenalenyl units ($\tau_1$) also has an angle dependence, which we model analogously to $t_a$ and $t_b$.

\section{Interacting theory and estimation of experimental twist angles}

\begin{figure}
 \centering
 \includegraphics[width=0.95\columnwidth]{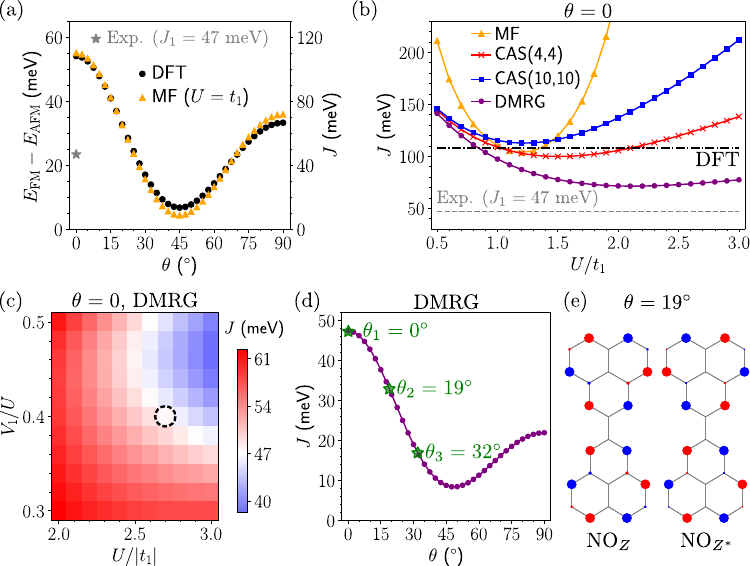}
 \caption{
 (a) Energy difference between triplet ferromagnetic and singlet antiferromagnetic DFT solutions of phenalenyl dimers, as a function of $\theta$ (black circles).
 Orange triangles denote results obtained within a MF Hubbard approximation, for $U=t_1$.
 Right axis labels the exchange coupling after spin decontamination, as given by Eq.~\eqref{eqJ_spin-decontaminated}.
 Gray star marks the experimental value of $J$ obtained by STM-IETS for $\theta=0$ (molecule \textbf{1} in Fig.~\ref{fig1}).
 (b) Exchange coupling of planar phenalenyl dimer, as a function of $U$, obtained with different methods to solve the effective Hubbard model.
 Black dash-dotted line marks the DFT result.
 Gray dashed line represents the experimental estimation.
 (c) Exchange coupling of planar phenalenyl dimer, as a function of $U$ and $V_1$, obtained by solving the nearest-neighbor extended Hubbard model via DMRG.
 Dashed circle indicates the parameters adopted in (d,e).
 (d) DMRG calculation of the exchange coupling of phenalenyl dimers, as a function of $\theta$.
 Green stars mark the estimated twist angles that match with the measured exchange couplings shown in Fig.~\ref{fig1}b.
 (e) Natural orbitals with occupation number close to 1, calculated by DMRG for the ground state of a phenalenyl dimer with $\theta=19^\circ$.
 Circle size scales with the modulus of the wave function, with blue and red indicating its sign.
 }
 \label{fig3}
\end{figure}

\subsection{Validation of Hubbard model and spin decontamination} 

As a first step toward elucidating the experimental spin excitations shown in Fig.~\ref{fig1}b, we carry out spin-polarized DFT calculations. 
We find that twisted phenalenyl dimers have a singlet antiferromagnetic ground state, followed by a triplet ferromagnetic first excited state.
In Fig.~\ref{fig3}a, we show the energy difference between the two magnetic configurations, as a function of $\theta$.
These results, complemented by the corresponding spin densities (SI-V), corroborate the established picture\cite{jacob_theory_2022,krane_exchange_2023,yu_magnetic_2023}---in line with Ovchinnikov-Lieb rules\cite{ovchinnikov_multiplicity_1978,lieb_two_1989}---that the low-energy properties of these molecules are well described by an antiferromagnetic spin-$\nicefrac{1}{2}$ dimer, where each effective spin ($\hat{\bm{S}}_\ell$, $\ell=1,2$) arises from single occupancy of the zero-energy wave function of the respective phenalenyl monomer.
Moreover, comparison with mean-field (MF) Hubbard model calculations (Fig.~\ref{fig3}a; see SI-V for details and additional results) shows that the effective tight-binding model derived in the previous section, together with electron-electron interactions in the form of a Hubbard repulsion $U$, is a reasonable starting point for describing the magnetic properties of this system.

Bearing in mind the Heisenberg model for a spin-$\nicefrac{1}{2}$ dimer,
\begin{equation}
    \hat{\mathcal{H}}=J\hat{\bm{S}}_1\cdot\hat{\bm{S}}_2,
\end{equation}
where $J>0$ is the antiferromagnetic exchange coupling, we define $J$ as the energy difference between the $S=0$ ground state and the $S=1$ first excited state.
However, in single-reference theories such as DFT or MF Hubbard, spin contamination implies that the energy of the antiferromagnetic solution (with total spin projection $S_z=0$) differs from that of the multi-reference $S=0$ ground state.
To demonstrate this, we adopt the low-energy picture mentioned above and write the ground state wave function of phenalenyl dimers as
\begin{equation}
    |\mathrm{GS}\rangle=\frac{1}{\sqrt{2}}(|\uparrow\rangle_{Z_1}|\downarrow\rangle_{Z_2}-|\downarrow\rangle_{Z_1}|\uparrow\rangle_{Z_2}),
    \label{eqGS}
\end{equation}
where $|\cdot\rangle_{Z_\ell}$ denotes the occupation of the zero-energy state of each phenalenyl monomer ($\ell=1,2$).
As for the first excited state, the triplet wave function reads
\begin{equation}
    |1\mathrm{ES}\rangle=
    \begin{cases}
      |\downarrow\rangle_{Z_1}|\downarrow\rangle_{Z_2},&S_z=-1\\
      \frac{1}{\sqrt{2}}(|\uparrow\rangle_{Z_1}|\downarrow\rangle_{Z_2}+|\downarrow\rangle_{Z_1}|\uparrow\rangle_{Z_2}),&S_z=0\\
      |\uparrow\rangle_{Z_1}|\uparrow\rangle_{Z_2},&S_z=+1
    \end{cases}.       
\end{equation}
Therefore, we see that while a single-determinant ferromagnetic solution 
can accurately describe the $S_z=\pm1$ multiplets of the first excited state, the same is not possible for the ground state.
Specifically, a single-reference theory will find an antiferromagnetic solution given by
\begin{equation}
    |\mathrm{AFM}\rangle=|\uparrow\rangle_{Z_1}|\downarrow\rangle_{Z_2}=\frac{1}{\sqrt{2}}(|\mathrm{GS}\rangle+|1\mathrm{ES},S_z=0\rangle).
\end{equation}
Using this insight, in Fig.~\ref{fig3}a we eliminate the spin contamination problem in the estimation of the exchange couplings by defining
\begin{equation}
    J=2(E_\mathrm{FM}-E_\mathrm{AFM}),
    \label{eqJ_spin-decontaminated}
\end{equation}
where $E_\mathrm{FM/AFM}$ denotes the total energy of the single-reference ferromagnetic/antiferromagnetic solution.

\subsection{Planar phenalenyl dimers: refining the theoretical description} 

We now focus on the results obtained for planar ($\theta=0$) phenalenyl dimers.
It is apparent that DFT calculations lead to an overestimation of $J$, by more than a factor of two, compared to experiments (Fig.~\ref{fig3}a).
One possible explanation for this disparity could be the absence of a substrate in our ab initio methodology, which may significantly alter the screening of electron-electron interactions.
Another possibility could be the need of employing DFT+U frameworks to accurately simulate carbon-based magnetic systems\cite{meena_ground-state_2022}.
As a crude approximation, these effects can be captured by adjusting the value of $U$ in a Hubbard model.
In Fig.~\ref{fig3}b, we show our MF Hubbard calculations as a function of $U$.
We observe that, regardless of $U$, the calculated $J$ remains largely overestimated.
Therefore, we conclude that theories such as MF Hubbard and DFT, which consider averaged interactions, are not suitable for quantitatively describing the exchange couplings in this system.

Following previous work\cite{ortiz_exchange_2019,jacob_theory_2022,biswas_steering_2023,krane_exchange_2023,henriques_designer_2024,turco_magnetic_2024,saleem_superexchange_2024}, we adopt an alternative approach wherein the effective Hubbard model is solved within a complete active space (CAS) approximation (see SI-VI for details).
This multi-reference method treats interactions to their full extent, but a truncation in the number of single-particle states used to span the many-body basis is required.
In Fig.~\ref{fig3}b, we present our results for two active spaces.
We observe that, for $U\lesssim~t_1$, our CAS calculations are stable with respect to the choice of the active space, but do not account for the experimental value of $J$.
In this limit, exchange interactions are dominated by Anderson's kinetic superexchange mechanisms\cite{anderson_new_1959,jacob_theory_2022} involving the molecular orbitals that host the effective spins (HOMO and LUMO), whose contribution is given by $J_\mathrm{kin}\simeq\frac{8|t_a-t_b|^2}{3U}$ for $U\gg|t_a-t_b|$.
As $U$ increases, Coulomb-driven superexchange mechanisms\cite{jacob_theory_2022}, of order $\mathcal{O}(U^2/t_1)$, become more relevant.
Their contributions, which can be ferromagnetic or antiferromagnetic, involve molecular orbitals beyond HOMO and LUMO, and are known to be unstable with the size of the active space\cite{jacob_theory_2022}.
This justifies the deviations observed between our CAS(4,4) and CAS(10,10) results for $U\gtrsim~t_1$ and casts doubt on the validity of the CAS approximation in this regime.

Given the limitations of the CAS method, we employ the density matrix renormalization group (DMRG)\cite{white_density_1992} to solve the effective Hubbard Hamiltonian.
Using the DMRG protocol outlined in SI-VII, the results obtained can be regarded as nearly exact solutions of the model.
As shown in Fig.~\ref{fig3}b, our DMRG estimates of $J$ are considerably closer to the experimental reference value for $U\sim2t_1$.
This suggests that the system may reside in a regime of strong interactions where the convergence of CAS calculations is challenging.
Therefore, we attest DMRG as a crucial method for accurately describing the magnetic interactions of phenalenyl dimers.

Although DMRG yields improved agreement with experiments, as shown in Fig.~\ref{fig3}b, the theoretical values of $J$ remain overestimated.
To address this discrepancy, we investigate the role of long-range interactions, inspired by prior studies\cite{jacob_theory_2022,krane_exchange_2023}.
Specifically, we augment the effective Hubbard Hamiltonian with a term that incorporates Coulomb repulsion $V_1$ between nearest-neighbor electrons,
\begin{equation}
    \hat{H}_{V_1}=\frac{V_1}{2}\sum_{{\langle}i,j\rangle}\hat{n}_i\hat{n}_j,
    \label{eqHV1}
\end{equation}
where $\hat{n}_i$ denotes the number operator for an electron at site $i$.
Considering that the repulsion between two electrons at sites $i$ and $j$, $V_{ij}$, is expected to decay inversely with distance according to Coulomb's law,
Eq.~\eqref{eqHV1} captures the dominant contribution of the general inter-site Coulomb Hamiltonian $\hat{H}_{V}=\frac{1}{2}\sum_{i{\neq}j}V_{ij}\hat{n}_i\hat{n}_j$.
The decision to focus solely on the dominant term is driven by two main factors.
First, a realistic determination of $V_{ij}$, including surface effects, poses a significant theoretical challenge that we defer to future work.
Consequently, we treat $V_1$ as a variable parameter, imposing an upper bound of $0.5U$, consistent with findings from gas-phase ab initio calculations\cite{jacob_theory_2022}.
Second, the simplified approach is also favored due to the efficiency of DMRG calculations for Hamiltonians with short-range interactions\cite{schollwock_density-matrix_2005,catarina_density-matrix_2023}.
In Fig.~\ref{fig3}c, we demonstrate that perfect agreement with experiments can be achieved within a physically plausible range of $(U,V_1)$ parameters.
We abstain from determining the exact point on the phase diagram corresponding to our h-BN/Rh(111) surface and simply adopt the parameters marked by the dashed circle: $U=2.7t_1$ and $V_1=0.4U$.
Additionally, it is worth noting that while our theory does not account for substrate-induced renormalizations of the spin excitation energies\cite{jacob_renormalization_2021}, these are expected to be small due to the presence of the h-BN decoupling layer.

\subsection{Angle-dependent exchange couplings} 

Using the model parameters derived above, we compute the exchange couplings as a function of $\theta$ (Fig.~\ref{fig3}d).
These calculations allow us to identify the twist angles compatible with the STM-IETS measurements shown in Fig.~\ref{fig1}b, thus supporting our experimental conjecture.
In particular, we attribute the different conductance step energies observed for the three phenalenyl dimers on different positions of the corrugated h-BN/Rh(111) surface to a geometric conformation in which the molecules exhibit dihedral angles of $\theta_1=0^{\circ}$ ($J_1=47~\si{\milli\electronvolt}$), $\theta_2=19^{\circ}$ ($J_2=33~\si{\milli\electronvolt}$), and $\theta_3=32^{\circ}$ ($J_3=17~\si{\milli\electronvolt}$).


\subsection{Natural orbitals and diradical character} 

To confirm the effective spin description of twisted phenalenyl dimers, we take the ($S=0$) ground state wave function obtained by DMRG for $\theta=\theta_2=19^{\circ}$ and compute the corresponding natural orbitals
(see SI-VIII for details and additional results).
We find two natural orbitals with occupancies near unity, NO$_Z$ and NO$_{Z^*}$ (Fig.~\ref{fig3}e), while the remaining feature occupation numbers close to 0 or 2.
Moreover, NO$_{Z/Z^*}$ closely resemble the bonding/antibonding combinations of the zero-energy states of each phenalenyl monomer.
These results, in alignment with CAS self-consistent field (CASSCF) calculations (SI-IX), suggest that the main contribution to the many-body wave function is an open-shell configuration akin to Eq.~\eqref{eqGS}.
Using DMRG, we have validated this diradical picture, obtaining $\sim71\%$ fidelity between the ground state and a many-body state where natural orbitals featuring occupancies near 0 (2) are considered empty (doubly-occupied), and the singly-occupied orbitals $\mathrm{NO}_{Z_{1/2}}=\frac{1}{\sqrt{2}}(\mathrm{NO}_{Z}\pm\mathrm{NO}_{Z^*})$ combine to form a singlet.

\section{Inducing twist angles via anchoring functional groups} 

\begin{figure}[t]
  \centering
  \includegraphics[width=0.95\columnwidth]{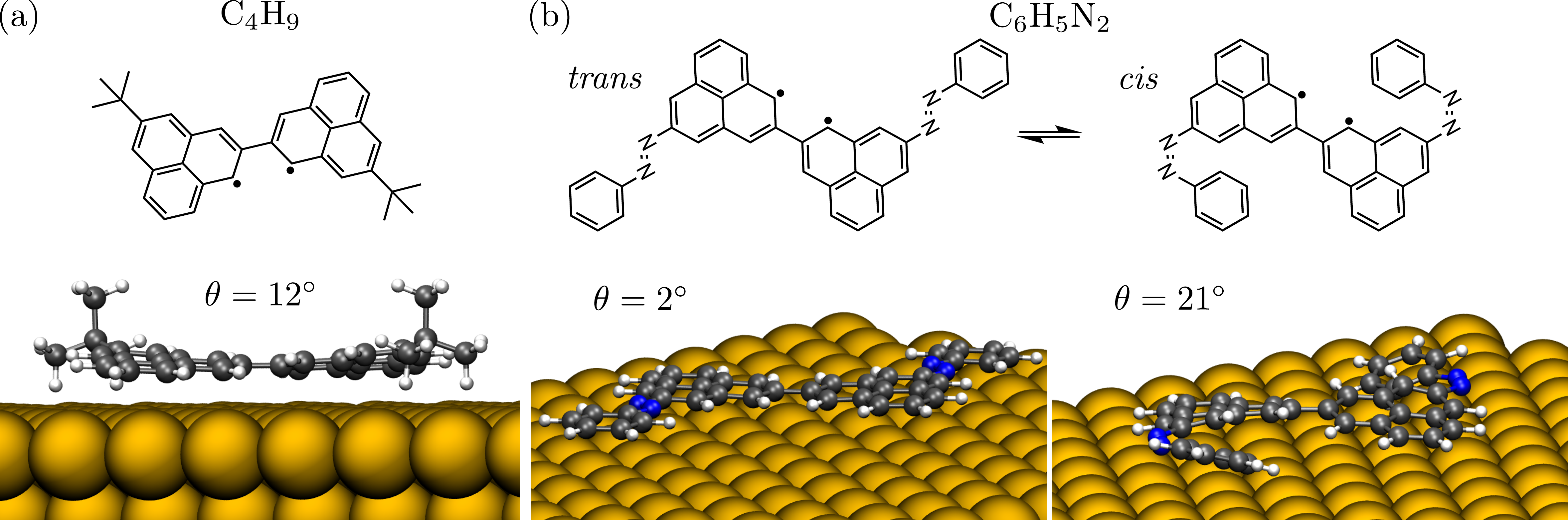}
  \caption{DFT geometry optimization of phenalenyl dimers on Au(111) with the following functional groups attached to two opposite edges: (a) tert-butyl (\ce{C4H9}), resulting in a dihedral angle $\theta=12^{\circ}$ between the phenalenyl molecules; (b) phenyldiazenyl (\ce{C6H5N2}) in \textit{trans} and \textit{cis} configurations, leading to $\theta=2^{\circ}$ and $\theta=21^{\circ}$, respectively.}
  \label{fig4}
\end{figure}

We now conduct first-principles geometry optimizations to investigate the possibility of inducing twist angles in phenalenyl dimers on Au(111) via anchoring functional groups.
We consider tert-butyl (\ce{C4H9}) and phenyldiazenyl (\ce{C6H5N2}) compounds covalently bonded to two opposite edges of the dimer.
Our results for tert-butyl (Fig.~\ref{fig4}a) show that the system lowers its energy by adopting a geometry with a dihedral angle of $\theta=12^{\circ}$ between phenalenyls.
For phenyldiazenyl (Fig.~\ref{fig4}b), we find that \textit{trans} and \textit{cis} isomers lead to $\theta=2^{\circ}$ and $\theta=21^{\circ}$, respectively.
Importantly, additional spin-polarized DFT calculations (SI-X) indicate that both functional groups do not significantly alter the magnetic properties of the dimer, aside from the $\theta$-dependent exchange couplings.

The previous results put forward a strategy for tuning the magnetic couplings in phenalenyl dimers through geometric conformations induced by functional groups.
The case of phenyldiazenyl is particularly noteworthy as it draws inspiration from azobenzene, whose \textit{cis} and \textit{trans} isomers are known to be photoswitchable\cite{henzl_reversible_2006}.
This opens the opportunity of dynamically tuning magnetic interactions with light, offering a novel approach to the design of light-responsive organic magnetic materials.

\section{Conclusion} 

Motivated by STM and STS measurements of twisted phenalenyl dimers on h-BN/Rh(111), we studied the twist-angle dependence of the exchange couplings in this diradical system.
Our theoretical investigation demonstrated that DMRG is a pivotal method for accurately describing the experimental results.
Based on our benchmarks, which we aim to extend to other systems in future work, we expect the methodology presented here to find widespread use in the modeling of open-shell nanographenes.
Furthermore, we proposed a strategy to reversibly switch the magnetic couplings of phenalenyl dimers with light, potentially enabling their use in quantum technology applications.

\begin{acknowledgement}

We thank Jo\~{a}o C. G. Henriques and Michal Jur\'{i}\v{c}ek for fruitful discussions.
This work was supported by the Swiss National Science Foundation (Grants No. CRSII5\_205987 and 200020\_212875), the NCCR MARVEL, a National Centre of Competence in Research funded by the Swiss National Science Foundation (Grant No. 205602), H2020-MSCA-ITN (ULTIMATE, No. 813036), and the Werner Siemens Foundation (CarboQuant). 
We acknowledge computational support from the Swiss Supercomputing Centre (CSCS) under the project ID s1267.
For the purpose of Open Access (which is required by our funding agencies), the authors have applied a CC BY public copyright license to any Author Accepted Manuscript version arising from this submission.

\end{acknowledgement}

\begin{suppinfo}

Experimental details, theoretical methods, and supplementary calculations, including additional Refs.~\citenum{catarina_hubbard_2022,
trambly_de_laissardiere_localization_2010,rozhkov_electronic_2016,
giannozzi_quantum_2009,perdew_generalized_1996,prandini2018precision,yakutovich_aiidalab_2021,pizzi2016aiida,hutter2014cp2k,vandevondele2007gaussian,goedecker1996separable,grimme2010consistent,hanke2013structure,
fishman_itensor_2022,schollwock_density-matrix_2011,
lain_density_2001,lowdin_quantum_1955,head-gordon_characterizing_2003,
orca,orca_5,dlpno_nevpt2,orca_def_jk}.

\end{suppinfo}

\providecommand{\latin}[1]{#1}
\makeatletter
\providecommand{\doi}
  {\begingroup\let\do\@makeother\dospecials
  \catcode`\{=1 \catcode`\}=2 \doi@aux}
\providecommand{\doi@aux}[1]{\endgroup\texttt{#1}}
\makeatother
\providecommand*\mcitethebibliography{\thebibliography}
\csname @ifundefined\endcsname{endmcitethebibliography}  {\let\endmcitethebibliography\endthebibliography}{}

\end{document}